\documentclass[letter,oldversion]{aa}  
\usepackage{graphicx}
\usepackage{natbib}
\usepackage{amsmath}
\usepackage{txfonts}
\usepackage{natbib}
\begin{document}

\title{Higher depletion of lithium in planet host stars: no age and mass effect}


   \author{S. G. Sousa\inst{1,}\inst{2}
    \and
    J. Fernandes\inst{3,}\inst{4}
    \and
    G. Israelian\inst{2,}\inst{5}
    \and
    N. C. Santos\inst{1}
     }
	  \institute{Centro de Astrof\'isica, Universidade do Porto, Rua das Estrelas, 4150-762 Porto, Portugal
	  \and  Instituto de Astrof\'isica de Canarias, 38200 La Laguna, Tenerife, Spain
	  \and Centro de F\'isica Computacional, Universidade de Coimbra, Coimbra, Portugal
	  \and Observat\'orio Astron\'omico e Departamento de Matem\'atica, Universidade de Coimbra, Coimbra, Portugal
	  \and Departamento de Astrofisica, Universidade de La Laguna, E-38205 La Laguna, Tenerife, Spain
     }
  
\date{--}
 
\abstract{Recent observational work by Israelian et al. has shown that sun-like planet host stars in the temperature range $5700$\,K$ < T_{eff} <5850$\,K have lithium abundances that are significantly lower than those observed for ``single'' field stars. In this letter we use stellar evolutionary models to show that differences in stellar mass and age are not responsible for the observed correlation. This result, along with the finding of Israelian et al., strongly suggest that the observed lithium difference is likely linked to some process related to the formation and evolution of planetary systems.
}
 
\authorrunning{Sousa et al.}

\maketitle

   \maketitle
%

\section{Introduction}

The increasing number of discovered planets continues to give new insights into the planet formation processes \citep[e.g.][]{Udry-2007b}. The study of planet host stars is providing crucial observational evidence for this. The first large uniform studies \citep[][]{Santos-2001,Santos-2004b} have shown that stars with giant planets have chemical abundances that are distinctly different from those found in ``single'' stars, a result that was later confirmed by other authors \citep[e.g.][]{Fischer_Valenti-2005}. This result provided important constraints for the models of planet formation and evolution \citep[e.g. ][]{Pollack-1996, Mordasini-2009, Boss-2002}.

Although most studies of chemical abundances in stars with planets have concentrated on the measurement of iron abundances as a metallicity proxy, several works have been made to study the abundances for a variety of refractory and volatile elements in planet host stars \citep[e.g.][]{Gilli-2006, Ecuvillon-2006b, Takeda-2007, Neves-2009}. 
Besides the general chemical enrichment found for stars hosting giant planets \citep[interestingly not found for stars hosting very low mass planets, see: ][]{Sousa-2008}, none of these found compelling evidence for other chemical peculiarities.

The possibility that stars with planets present different abundances of the light elements lithium and beryllium has also been debated \citep[][]{King-1997, GarciaLopez-1998, Deliyannis-2000, Cochran-1997, Ryan-2000, Gonzalez-2000, Gonzalez-2008, Israelian-2004, Santos-2004c, Takeda-2005, Takeda-2007, Chen-2006, Luck-2006}. Different abundances could indicate that planets or planetary material were engulfed by the star during its lifetime (and in which quantity)  \citep[e.g.][]{Israelian-2001}, or suggest that the rotational history of the stars depends on the existence of planets or indirectly on the process for planet formation \citep[][]{Bouvier-2008, Castro-2009}.

Recently, \citet[][]{Israelian-2009} presented a large uniform study of lithium abundances in a sample of stars from the HARPS planet search programme \citep[][]{Mayor-2003}. In their work they reported a significant difference regarding the depletion of lithium for planet host stars when compared with stars with no detected planet in the range $5700$\,K$ < T_{eff} <5850$\,K, confirming former suspicions \citep[e.g.][]{Israelian-2004,Takeda-2007}. According to Israelian et al., stars with planets in the temperature range around the solar temperature (solar analogues) have significantly lower lithium abundances when compared with ``single'' stars (for which no planets were detected so far). The uniformity of the HARPS sample (composed mainly of old inactive stars) allowed Israelian et al. to exclude effects like stellar rotation, stellar activity, or chemical abundances as the cause for the observed difference.

In this paper we use stellar evolution models to explore the possibility that stellar age and mass could be responsible for the observed difference. In Sect.\,2 we present the procedure used for the determination of precise and uniform masses and ages for our sample. In Sect.\,3 we explore possible correlations between these two parameters and the lithium abundances. We conclude in Sect.\,4.

\section{Masses and ages}

The stellar age and mass was determined by means of comparison between stellar evolutionary models and observations. Individual values of the stellar luminosity, effective temperature, and metallicity for the solar analogues in our sample \citep[][]{Israelian-2009} were taken from the uniform study of \citep[][]{Sousa-2008}. These were used for a comparison with stellar models computed with the CESAM code version 3 \citep[][]{Morel-1997}, running in the Coimbra Observatory. The details on the physics of these models can be found in \citet[][]{Fernandes-2004} with one exception: the equation of state EFF is used \citep[][]{Eggleton-1973}, an analytical equation of state suitable for FGK stars. With these input physics the solar model fits the observed luminosity and effective temperature for the common accepted solar age of 4.6~Gyr \citep[][]{Dziembowski-1999}, as better as $10^{-4}$, considering a mixing length parameter of convection $\alpha{=}1.65$, the helium abundance $Y{=}0.2675$, and the metal abundance $Z{=}0.0173$. With these values, and assuming the primordial helium equal to 0.25 \citep[e.g.][]{Peimbert-2009}, the helium abundance to metallicity ratio is $\Delta~Y/\Delta~Z=1.53$. The mass and age resolution for the models used in this procedure are 0.01 $M_{\sun}$ and a maximum time step of 200 Myr respectively.

For each star with a metal abundance fixed to the observed metallicity value \citep[according to][]{Sousa-2008}, we computed several models for different ages and masses to fit the observed luminosity and effective temperature. We assumed for those models the solar mixing length parameter, a null overshooting parameter ($\alpha_{ov}$) for stars with masses lower than 1.1 $M_{\sun}$, and $\alpha_{ov}=0.2$ for higher mass stars \citep[e.g.][]{Claret-2007}. 

The stellar mass and age are estimated as follows: for each stellar evolutionary model we compute the functional 
\[
 \chi^2 ( M, t) = \left( \frac{\log L_{obs} - \log L_{mod}}{\sigma_L} \right)^2 + \left( \frac{\log Teff_{obs} - \log Teff_{mod}}{\sigma_{Teff}} \right)^2
\]
where the subscripts $obs$ and $mod$ refer to the observed values and theoretical (model) values respectively \citep[e.g.][]{Lastennet-2002}. The $\sigma$ values are taken from \citet[][]{Sousa-2008}. The solution is obtained by minimizing the above function. For each star the final mass and age is obtained when the prediction of the theoretical stellar model is inside the observational error bar, i.e., $\chi^2 (M,t) < 2$.

In Table \ref{tab1} we present the derived values for mass and age in our sample. The typical (relative) errors bars of our estimations are 0.01$M_{\sun}$ in mass and 0.5 Gyr and 1.0 Gyr, respectively, for ages lower and higher than 5 Gyr. This level of precision is possible because of the very small (relative) errors in the stellar parameters derived for the solar analogues in \citet[][]{Sousa-2008}. We note however that we are not taking into account variations in parameters like the mixing length and helium abundance. Changes in these parameters could lead to errors in the stellar masses and ages \citep[e.g.][]{Fernandes-2004}.


\begin{figure}[t!]
\centering
\includegraphics[width=8cm]{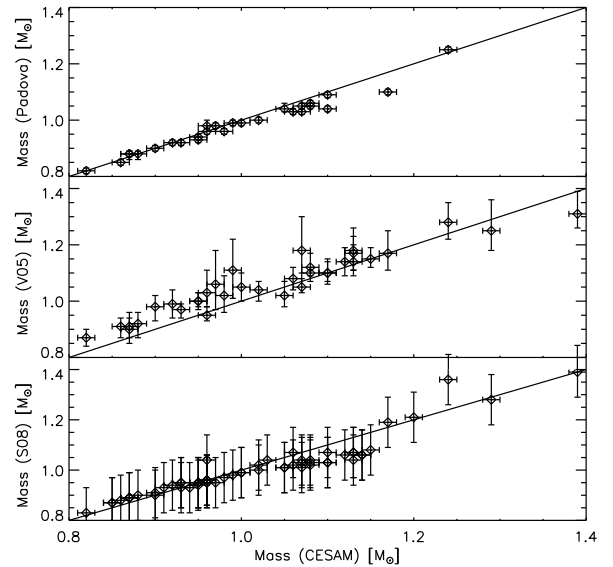}
\caption[]{Comparison between masses determined in this work using CESAM with the ones presented in \citet[][]{Sousa-2008} (bottom), \citet[][]{Valenti-2005} (midle) and determined using the Padova models (top). The filled lines represent the identity line.}
\label{masses}
\end{figure}

\begin{figure}[t!]
\centering
\includegraphics[width=8cm]{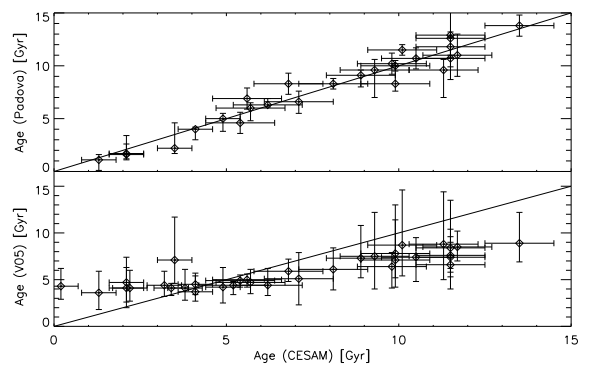}
\caption[]{Same as Fig.\ref{masses}, but for ages.}
\label{ages}
\end{figure}

In order to check our methodology, we applied the stellar evolutionary models from the Padova group, computed with the web interface dealing with stellar isochrones and their derivatives (http://stev.oapd.inaf.it/cgi-bin/cmd) to the stars of our sample. Taking into account that the Padova models are constrained to metal abundance lower than Z $<$ 0.03, we have arbitrarily chosen a subsample of 27 stars to cover the range of mass and age of the stars in the paper. In Figs. \ref{masses} and \ref{ages} we show the comparison between masses and ages respectively, using the CESAM and Padova models. In addition we also include a comparison to the values published in \citet[][]{Sousa-2008} and \citet[][]{Valenti-2005} (with this latter we have 41 stars in common). A very good agreement exists between all sets of values, except for the ages of \citet[][]{Valenti-2005} which seem to be constrained to values closer to the Sun age. The reason for this is not clear; this discussion is out of the scope of the present paper. However, if the age span of the stars in our sample were smaller (taking the values from Valenti \& Fischer), it would strengthen the results discussed below.

\begin{table*}[!t]
\centering 
\caption[]{New mass and age values together with published values for lithium abundance ($<$ represents upper limit), [Fe/H] and Teff.}
\begin{tabular}{|lccccc|lccccc|}
\hline
\hline
\noalign{\smallskip}
star(HD)& mass($M_{\sun}$) & Age(Gy) & $\log[N(Li)]$ & [Fe/H] & Teff & star(HD) & mass($M_{\sun}$) & Age(Gy)  & $\log[N(Li)]$ & [Fe/H] & Teff \\
\hline
\textbf{planet hosts:} &   &    &  &   &   &      45289 & 0.95 & 11.5 & $<$0.47 & -0.02  & 5717  \\
 16141 & 1.14 &  6.0 & $<$1.00  &  0.16 & 5806 &  76151 & 1.07 &  1.3 & 1.87    &  0.12  & 5788  \\
 16417 & 1.20 &  5.2 & 1.85     &  0.13 & 5841 &  78429 & 1.02 &  8.1 & $<$0.35 &  0.09  & 5760  \\
 20782 & 0.95 &  9.9 & $<$0.47  & -0.06 & 5774 &  78538 & 1.02 &  1.3 & 2.44    & -0.03  & 5786  \\
 66428 & 1.07 &  5.7 & $<$0.71  &  0.25 & 5705 &  78612 & 0.93 & 11.8 & 1.62    & -0.24  & 5834  \\
 92788 & 1.10 &  2.2 & $<$0.82  &  0.27 & 5744 &  88084 & 0.94 &  7.9 & $<$0.91 & -0.10  & 5766  \\
107148 & 1.13 &  3.2 & $<$1.24  &  0.31 & 5805 &  89454 & 1.07 & $<$0.5 & 1.64  &  0.12 & 5728  \\
114729 & 0.93 & 11.6 & 2.00     & -0.28 & 5844 &  92719 & 1.00 &  1.7 & 1.93    & -0.10  & 5824  \\
134987 & 1.08 &  6.2 & $<$0.60  &  0.25 & 5740 &  95521 & 0.95 &  4.1 & 1.67    & -0.15  & 5773  \\
160691 & 1.15 &  5.2 & $<$0.98  &  0.30 & 5780 &  96700 & 0.92 & 11.5 & 1.35    & -0.18  & 5845  \\
202206 & 1.13 & $<$0.5 & 1.45   &  0.29 & 5757 &  97998 & 0.82 & 10.1 & 1.72    & -0.42  & 5716  \\
204313 & 1.08 &  2.9 & $<$0.52  &  0.18 & 5776 & 108309 & 1.10 &  6.8 & $<$0.94 &  0.12 &  5775  \\
222582 & 0.98 &  8.9 & $<$0.91  & -0.01 & 5779 & 111031 & 1.12 &  3.8 & $<$0.75 &  0.27 &  5801  \\
\multicolumn{2}{|l}{\textbf{comparison stars:}} & & & & & 114613 & 1.24 &  5.4  & 2.69  &  0.19 &  5729  \\
  1461 & 1.06 &  4.9 & $<$0.74  &  0.19 & 5765 & 114853 & 0.86 & 11.5 & $<$0.46 & -0.23 &  5705  \\
  2071 & 0.96 &  3.5 & $<$1.43  & -0.09 & 5719 & 115585 & 1.14 &  5.8 & $<$0.51 &  0.35 &  5711  \\
  4307 & 0.99 &  9.8 & 2.48     & -0.23 & 5812 & 145809 & 0.97 & 10.5 & 2.13    & -0.25 &  5778  \\
  8406 & 0.96 &  3.3 & 1.72     & -0.10 & 5726 & 146233 & 1.05 &  2.1 & 1.64    &  0.04 &  5818  \\
 11505 & 0.87 & $>$14 & $<$0.35 & -0.22 & 5752 & 154962 & 1.29 &  4.1 & 2.39    &  0.32 &  5827  \\
 12387 & 0.85 & $>14$ & $<$0.15 & -0.24 & 5700 & 183658 & 1.00 &  7.1 & $<$1.09 &  0.03 &  5803  \\
 19467 & 0.90 & 13.5 & $<$0.55  & -0.14 & 5720 & 189567 & 0.85 & 13.9 & $<$0.18 & -0.24 &  5726  \\
 20619 & 0.87 &  9.9 & 1.71     & -0.22 & 5703 & 189625 & 1.10 &  2.1 & 2.12    &  0.18 &  5846  \\
 21938 & 0.78 & $>$14 & $<$1.13 & -0.47 & 5778 & 198075 & 0.96 &  1.5 & 1.95    & -0.24 &  5846  \\
 27063 & 1.06 & $<$0.5 & 1.70   & 0.05  & 5767 & 208704 & 0.96 &  9.5 & $<$1.09 & -0.09  & 5826  \\
 28471 & 0.95 &  8.7 & $<$0.73  & -0.05 & 5745 & 210918 & 0.93 & 11.7 & $<$0.28 & -0.09  & 5755  \\
 32724 & 0.96 & 11.0 & 1.63     & -0.17 & 5818 & 211415 & 0.91 &  9.7 & 1.85    & -0.21  & 5850  \\
 34449 & 1.03 & $<$0.5 & 2.08   & -0.09 & 5848 & 215456 & 1.05 &  8.5 & 2.38    & -0.09  & 5789  \\
 37962 & 0.87 & 11.3 & 1.84     & -0.20 & 5718 & 220507 & 0.96 & 11.5 & $<$0.56 &  0.01  & 5698  \\
 38858 & 0.88 &  9.3 & 1.54     & -0.22 & 5733 & 221420 & 1.39 &  3.4 & 2.75    &  0.33  & 5847  \\
 44420 & 1.13 &  2.2 & $<$0.71  &  0.29 & 5818 & 223171 & 1.17 &  5.6 & 2.11    &  0.12  & 5841  \\
 44594 & 1.08 &  4.1 & 1.55     &  0.15 & 5840 & 224393 & 0.90 &  2.1 & 2.25    & -0.38  & 5774  \\

\hline
\end{tabular}
\label{tab1}
\end{table*}

\section{Lithium vs. age}

Table\,\ref{tab1} shows the values of mass and age derived for the stars in this work and their lithium abundances \citep[][]{Israelian-2009}. The sample studied here corresponds to the majority of solar analogues studied in \citet[][]{Israelian-2009}. However, here we only considered those belonging to the HARPS sample ($\sim$60). For these stars we have the best uniform stellar parameters \citep[][]{Sousa-2008}, assuring the total consistency of our results.
Four stars in the sample were also excluded since no reliable masses and ages could be derived (HD\,28701, HD\,78558, HD\,96423, HD\,143114).

\begin{figure}[t!]
\centering
\includegraphics[width=8cm]{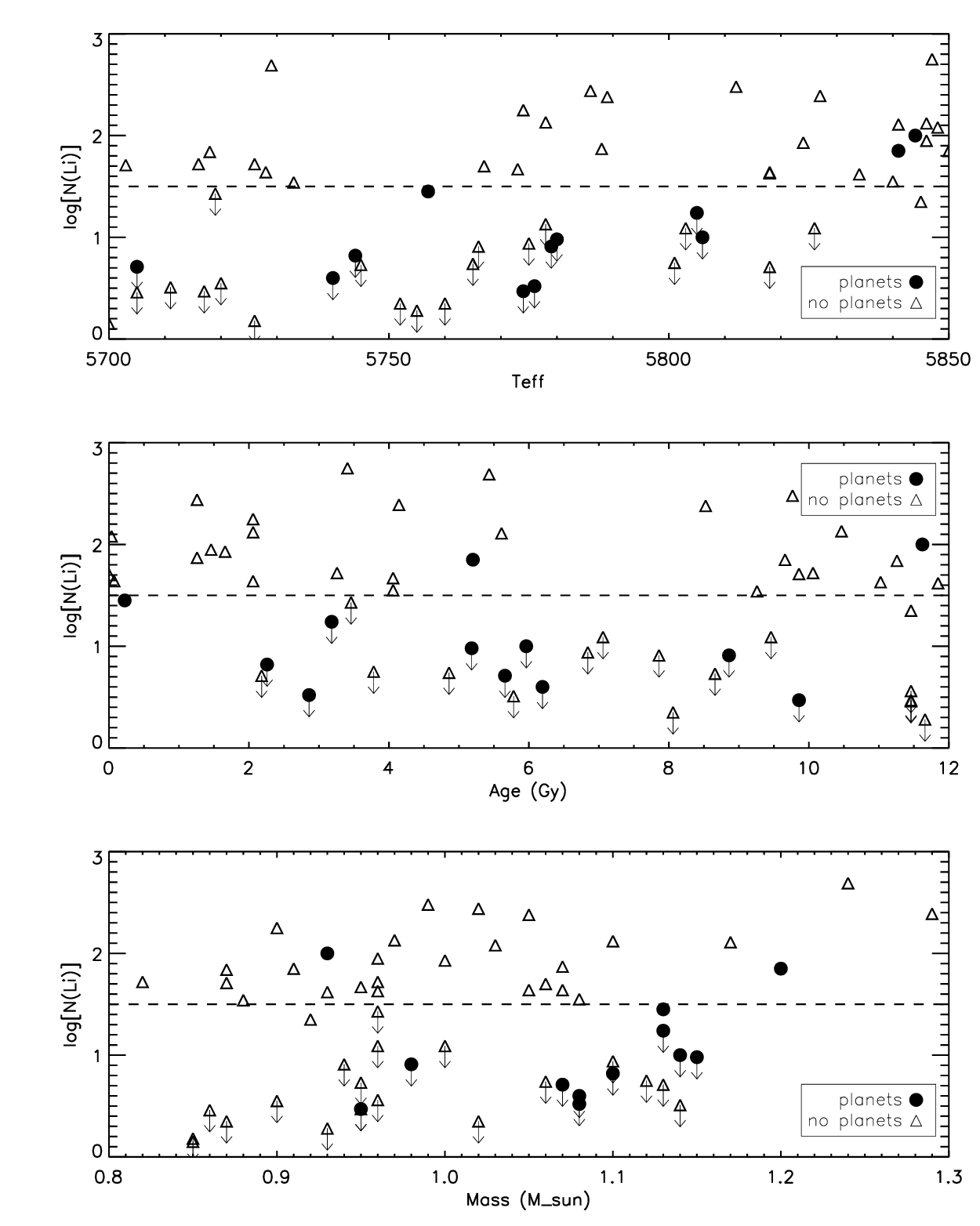}
\caption[]{Lithium abundances as a function of effective temperature (top), age (middle), and mass (lower panel). Filled circles denote stars with planets, while open triangles denote ''single'' stars. Arrows indicate upper limits.}
\label{fig_calibration_proc}
\end{figure}
The top panel in Fig. \ref{fig_calibration_proc} shows the lithium abundance plotted against the effective temperature for the stars presented in Table \ref{tab1}. This result is almost indistinguishable from Fig.\,1 in \citet[][]{Israelian-2009}. There is a general tendency for lower temperature stars to present on average lower lithium abundances, an expected result \citep[see e.g.][]{Pinsonneault-1990, Sestito-2005}. It is also evident that the stars that host planets are distributed in the lower region of the plot, while the stars with no evidence for planetary companions are spread over the entire diagram.

\begin{figure}[t!]
\centering
\includegraphics[width=8cm]{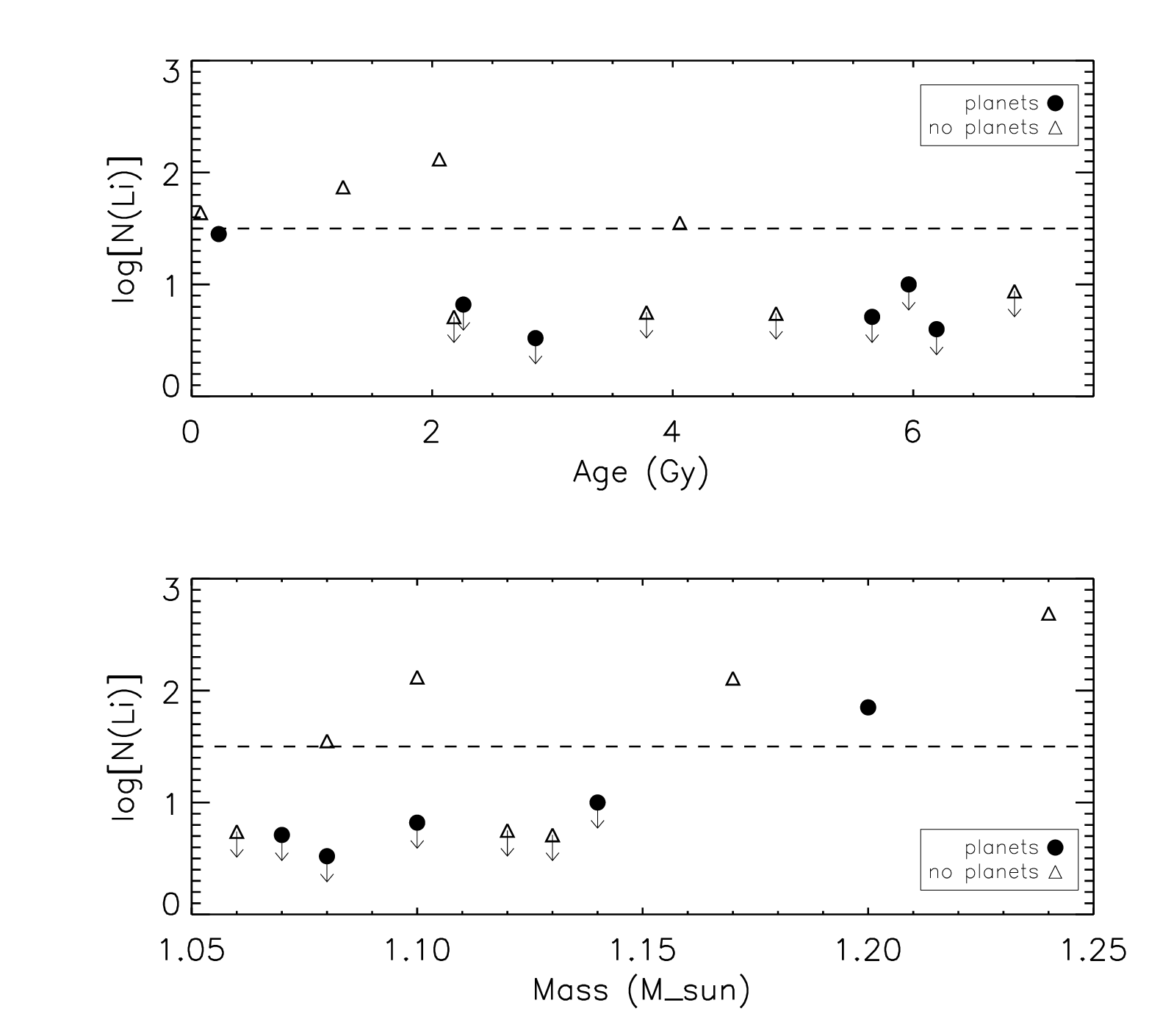}
\caption[]{Lithium abundances as a function of stellar mass (top) and age (bottom) in restricted subsamples of our stars.}
\label{fig3}
\end{figure}
In the middle plot of Fig. \ref{fig_calibration_proc} we present the lithium abundance as a function of the derived stellar age. As expected, the bulk of the stars are old. The HARPS sample was chosen to avoid active (younger) stars due to the higher difficulty in searching for planets in these objects \citep[e.g.][]{Santos-2000a}. 
Interestingly a few younger objects are seen in the left part of this panel ($<$2\,Gyr), a region where no lithium-depleted star is seen. This suggests a lithium abundance dependence with age. Above an age of $\sim$2\,Gyr, however, we observe a wider dispersion in lithium for all stars in the plot, and no correlation is seen.

The middle plot of Fig. \,2 also clearly shows that even though we have planet hosts stars in all the age interval, these are preferably located at lower lithium abundances. No particular age correlation is present. Age does not seem to explain the observed difference between the lithium abundances in the two samples.

Finally, we present in the bottom panel the lithium abundance as a function the derived stellar mass. As for the age, we can easily see that the dispersion of the lithium abundances has the same behaviour for the complete range of masses in the sample. No evidence is found that stellar mass could be responsible for the observed lithium offset between the two samples.
We note that the two planet hosts with the highest lithium abundances show no particular age or mass tendency. Their high abundance is likely explained because they are among the highest temperature stars in our sample.

We have also verified if the relatively large range in metallicities for our stars together with the dispersion in ages and masses could not be hiding any trend. To do this, we selected only stars within the metallicity interval 0.1$<$[Fe/H]$<$0.3, where the largest overlap between stars with and without planets exists. From these we selected only those with masses between 1.05 and 1.15\,M$_\odot$ (Fig.\,3, upper panel, lithium as a function of stellar age), and with ages between 2\,Gyr and 6\,Gyr (Fig.\,3, lower panel, lithium as a function of stellar mass). The plots of Fig.\ref{fig3} do not show any hint of a mass or age-lithium correlation that could explain the observed difference between planet host and ``single'' stars.

\section{Conclusions}

We derived uniform values for the stellar ages and masses in a sample of solar analogue stars, for which \citet[][]{Israelian-2009} have found a clear difference in the lithium abundances correlated to the existence of planets. Our analysis has shown that age and mass cannot explain the lithium abundance differences. This strongly suggests that the observed differences are not related to stellar intrinsic properties. This result confirms the uniformity of the studied sample.
  
Our results are directly linked with Li observations in open clusters. A large high-quality database for Li in open clusters acquired in the last years has permitted us to draw a secure picture of the empirical behaviour and evolution of Li in solar analogue stars \citep[][]{Randich-2008, Randich-2009, Sestito-2005}. Numerous observations show that stars with the same age, temperature, and metallicity can be affected by different amounts of Li depletion. A dispersion in Li of at least a factor of 10 has been observed in the solar-age, solar metallicity clusters M 67 \citep[][]{Pasquini-2008} and several other old, solar metallicity or metal rich clusters \citep[][]{Randich-2008, Randich-2009}. This spread strongly suggests that Li depletion must be affected by an additional parameter besides mass, age, and chemical composition. This parameter could be an initial angular momentum of the star affected by the formation and evolution of planets. Similar Li dispersion (or often cited as bi-modality) is also observed in solar type field stars \citep[][]{Favata-1996, Galeev-2004, Chen-2001, Lambert-2004}. We think that the same mechanism is responsible for the large Li dispersion in the field and open cluster solar type stars. Figure \ref{fig_calibration_proc} clearly supports this suggestion.

No clear explanation has been found for the lithium abundance difference observed between stars with and without planets, though a few possibilities have been suggested. These include the star-planet interaction \citep[][]{Castro-2009}, the infall of planets into the star (leading to higher mixing - Theado et al., priv. comm.), or a difference in the rotational history of the star due to star-disc interaction. Massive proto-planetary discs capable of forming planets will likely help to break stellar rotation, thus changing the depletion rate of lithium \citep[e.g.][]{Charbonnel-2005, Cochran-1997, Bouvier-2008, Pinsonneault-2010}. Interestingly, the same mechanism could have been responsible for the low lithium abundance observed in our own Sun, itself a planet host stars, and can be used to explain the dispersion of lithium abundances observed in clusters of different ages \citep[][]{Sestito-2005}.

\begin{acknowledgements}
S.G.S acknowledges the support from the Funda\c{c}\~ao para a Ci\^encia e Tecnologia (Portugal) in the form of a grants SFRH/BPD/47611/2008. NCS thanks for the support by the European Research Council/European Community under the FP7 through a Starting Grant, as well as the support from Funda\c{c}\~ao para a Ci\^encia e a Tecnologia (FCT), Portugal, through programme Ci\^encia\,2007. We also acknowledge support from FCT in the form of grants reference PTDC/CTE-AST/098528/2008 and PTDC/CTE-AST/66181/2006. This research has also been supported by the Spanish Ministry of Science and Innovation (MICINN, AYA2008-04874). Finally we thank the referee for its useful comments.
\end{acknowledgements}

\bibliographystyle{aa}
\bibliography{sousa_bibliography}

\begin{thebibliography}{49}
\expandafter\ifx\csname natexlab\endcsname\relax\def\natexlab#1{#1}\fi

\bibitem[{{Boss}(2002)}]{Boss-2002}
{Boss}, A.~P. 2002, \apjl, 567, L149

\bibitem[{{Bouvier}(2008)}]{Bouvier-2008}
{Bouvier}, J. 2008, \aap, 489, L53

\bibitem[{{Castro} {et~al.}(2009){Castro}, {Vauclair}, {Richard}, \&
  {Santos}}]{Castro-2009}
{Castro}, M., {Vauclair}, S., {Richard}, O., \& {Santos}, N.~C. 2009, \aap,
  494, 663

\bibitem[{{Charbonnel} \& {Primas}(2005)}]{Charbonnel-2005}
{Charbonnel}, C. \& {Primas}, F. 2005, \aap, 442, 961

\bibitem[{{Chen} {et~al.}(2001){Chen}, {Nissen}, {Benoni}, \&
  {Zhao}}]{Chen-2001}
{Chen}, Y.~Q., {Nissen}, P.~E., {Benoni}, T., \& {Zhao}, G. 2001, \aap, 371,
  943

\bibitem[{{Chen} \& {Zhao}(2006)}]{Chen-2006}
{Chen}, Y.~Q. \& {Zhao}, G. 2006, \aj, 131, 1816

\bibitem[{{Claret}(2007)}]{Claret-2007}
{Claret}, A. 2007, \aap, 475, 1019

\bibitem[{{Cochran} {et~al.}(1997){Cochran}, {Hatzes}, {Butler}, \&
  {Marcy}}]{Cochran-1997}
{Cochran}, W.~D., {Hatzes}, A.~P., {Butler}, R.~P., \& {Marcy}, G.~W. 1997,
  \apj, 483, 457

\bibitem[{{Deliyannis} {et~al.}(2000){Deliyannis}, {Cunha}, {King}, \&
  {Boesgaard}}]{Deliyannis-2000}
{Deliyannis}, C.~P., {Cunha}, K., {King}, J.~R., \& {Boesgaard}, A.~M. 2000,
  \aj, 119, 2437

\bibitem[{{Dziembowski} {et~al.}(1999){Dziembowski}, {Fiorentini}, {Ricci}, \&
  {Sienkiewicz}}]{Dziembowski-1999}
{Dziembowski}, W.~A., {Fiorentini}, G., {Ricci}, B., \& {Sienkiewicz}, R. 1999,
  \aap, 343, 990

\bibitem[{{Ecuvillon} {et~al.}(2006){Ecuvillon}, {Israelian}, {Santos},
  {Mayor}, \& {Gilli}}]{Ecuvillon-2006b}
{Ecuvillon}, A., {Israelian}, G., {Santos}, N.~C., {Mayor}, M., \& {Gilli}, G.
  2006, A\&A, 449, 809

\bibitem[{{Eggleton} {et~al.}(1973){Eggleton}, {Faulkner}, \&
  {Flannery}}]{Eggleton-1973}
{Eggleton}, P.~P., {Faulkner}, J., \& {Flannery}, B.~P. 1973, \aap, 23, 325

\bibitem[{{Favata} {et~al.}(1996){Favata}, {Micela}, \&
  {Sciortino}}]{Favata-1996}
{Favata}, F., {Micela}, G., \& {Sciortino}, S. 1996, \aap, 311, 951

\bibitem[{{Fernandes} \& {Santos}(2004)}]{Fernandes-2004}
{Fernandes}, J. \& {Santos}, N.~C. 2004, \aap, 427, 607

\bibitem[{{Fischer} \& {Valenti}(2005)}]{Fischer_Valenti-2005}
{Fischer}, D.~A. \& {Valenti}, J. 2005, \apj, 622, 1102

\bibitem[{{Galeev} {et~al.}(2004){Galeev}, {Bikmaev}, {Mashonkina}, {Musaev},
  \& {Galazutdinov}}]{Galeev-2004}
{Galeev}, A.~I., {Bikmaev}, I.~F., {Mashonkina}, L.~I., {Musaev}, F.~A., \&
  {Galazutdinov}, G.~A. 2004, Astronomy Reports, 48, 511

\bibitem[{{Garcia Lopez} \& {Perez de Taoro}(1998)}]{GarciaLopez-1998}
{Garcia Lopez}, R.~J. \& {Perez de Taoro}, M.~R. 1998, \aap, 334, 599

\bibitem[{{Gilli} {et~al.}(2006){Gilli}, {Israelian}, {Ecuvillon}, {Santos}, \&
  {Mayor}}]{Gilli-2006}
{Gilli}, G., {Israelian}, G., {Ecuvillon}, A., {Santos}, N.~C., \& {Mayor}, M.
  2006, A\&A, 449, 723

\bibitem[{{Gonzalez}(2008)}]{Gonzalez-2008}
{Gonzalez}, G. 2008, \mnras, 386, 928

\bibitem[{{Gonzalez} \& {Laws}(2000)}]{Gonzalez-2000}
{Gonzalez}, G. \& {Laws}, C. 2000, AJ, 119, 390

\bibitem[{{Israelian} {et~al.}(2009){Israelian}, {Delgado Mena}, {Santos},
  {Sousa}, {Mayor}, {Udry}, {Dom{\'{\i}}nguez Cerde{\~n}a}, {Rebolo}, \&
  {Randich}}]{Israelian-2009}
{Israelian}, G., {Delgado Mena}, E., {Santos}, N.~C., {et~al.} 2009, \nat, 462,
  189

\bibitem[{{Israelian} {et~al.}(2001){Israelian}, {Santos}, {Mayor}, \&
  {Rebolo}}]{Israelian-2001}
{Israelian}, G., {Santos}, N.~C., {Mayor}, M., \& {Rebolo}, R. 2001, Nature,
  411, 163

\bibitem[{{Israelian} {et~al.}(2004){Israelian}, {Santos}, {Mayor}, \&
  {Rebolo}}]{Israelian-2004}
{Israelian}, G., {Santos}, N.~C., {Mayor}, M., \& {Rebolo}, R. 2004, \aap, 414,
  601

\bibitem[{{King} {et~al.}(1997){King}, {Deliyannis}, {Hiltgen}, {Stephens},
  {Cunha}, \& {Boesgaard}}]{King-1997}
{King}, J.~R., {Deliyannis}, C.~P., {Hiltgen}, D.~D., {et~al.} 1997, \aj, 113,
  1871

\bibitem[{{Lambert} \& {Reddy}(2004)}]{Lambert-2004}
{Lambert}, D.~L. \& {Reddy}, B.~E. 2004, \mnras, 349, 757

\bibitem[{{Lastennet} \& {Valls-Gabaud}(2002)}]{Lastennet-2002}
{Lastennet}, E. \& {Valls-Gabaud}, D. 2002, \aap, 396, 551

\bibitem[{{Luck} \& {Heiter}(2006)}]{Luck-2006}
{Luck}, R.~E. \& {Heiter}, U. 2006, \aj, 131, 3069

\bibitem[{{Mayor} {et~al.}(2003){Mayor}, {Pepe}, {Queloz}, {Bouchy},
  {Rupprecht}, {Lo Curto}, {Avila}, {Benz}, {Bertaux}, {Bonfils}, {dall},
  {Dekker}, {Delabre}, {Eckert}, {Fleury}, {Gilliotte}, {Gojak}, {Guzman},
  {Kohler}, {Lizon}, {Longinotti}, {Lovis}, {Megevand}, {Pasquini}, {Reyes},
  {Sivan}, {Sosnowska}, {Soto}, {Udry}, {van Kesteren}, {Weber}, \&
  {Weilenmann}}]{Mayor-2003}
{Mayor}, M., {Pepe}, F., {Queloz}, D., {et~al.} 2003, The Messenger, 114, 20

\bibitem[{{Mordasini} {et~al.}(2009){Mordasini}, {Alibert}, \&
  {Benz}}]{Mordasini-2009}
{Mordasini}, C., {Alibert}, Y., \& {Benz}, W. 2009, \aap, 501, 1139

\bibitem[{{Morel}(1997)}]{Morel-1997}
{Morel}, P. 1997, \aaps, 124, 597

\bibitem[{{Neves} {et~al.}(2009){Neves}, {Santos}, {Sousa}, {Correia}, \&
  {Israelian}}]{Neves-2009}
{Neves}, V., {Santos}, N.~C., {Sousa}, S.~G., {Correia}, A.~C.~M., \&
  {Israelian}, G. 2009, \aap, 497, 563

\bibitem[{{Pasquini} {et~al.}(2008){Pasquini}, {Biazzo}, {Bonifacio},
  {Randich}, \& {Bedin}}]{Pasquini-2008}
{Pasquini}, L., {Biazzo}, K., {Bonifacio}, P., {Randich}, S., \& {Bedin}, L.~R.
  2008, \aap, 489, 677

\bibitem[{{Peimbert} {et~al.}(2009){Peimbert}, {Peimbert}, {Carigi}, \&
  {Luridiana}}]{Peimbert-2009}
{Peimbert}, M., {Peimbert}, A., {Carigi}, L., \& {Luridiana}, V. 2009, ArXiv
  e-prints

\bibitem[{{Pinsonneault}(2010)}]{Pinsonneault-2010}
{Pinsonneault}, M.~H. 2010, ArXiv e-prints

\bibitem[{{Pinsonneault} {et~al.}(1990){Pinsonneault}, {Kawaler}, \&
  {Demarque}}]{Pinsonneault-1990}
{Pinsonneault}, M.~H., {Kawaler}, S.~D., \& {Demarque}, P. 1990, \apjs, 74, 501

\bibitem[{{Pollack} {et~al.}(1996){Pollack}, {Hubickyj}, {Bodenheimer},
  {Lissauer}, {Podolak}, \& {Greenzweig}}]{Pollack-1996}
{Pollack}, J.~B., {Hubickyj}, O., {Bodenheimer}, P., {et~al.} 1996, Icarus,
  124, 62

\bibitem[{{Randich}(2008)}]{Randich-2008}
{Randich}, S. 2008, Memorie della Societa Astronomica Italiana, 79, 516

\bibitem[{{Randich} {et~al.}(2009){Randich}, {Pace}, {Pastori}, \&
  {Bragaglia}}]{Randich-2009}
{Randich}, S., {Pace}, G., {Pastori}, L., \& {Bragaglia}, A. 2009, \aap, 496,
  441

\bibitem[{{Ryan}(2000)}]{Ryan-2000}
{Ryan}, S.~G. 2000, \mnras, 316, L35

\bibitem[{{Santos} {et~al.}(2004{\natexlab{a}}){Santos}, {Israelian},
  {Garc{\'{\i}}a L{\' o}pez}, {Mayor}, {Rebolo}, {Randich}, {Ecuvillon}, \&
  {Dom{\'{\i}}nguez Cerde{\~ n}a}}]{Santos-2004c}
{Santos}, N.~C., {Israelian}, G., {Garc{\'{\i}}a L{\' o}pez}, R.~J., {et~al.}
  2004{\natexlab{a}}, A\&A, 427, 1085

\bibitem[{{Santos} {et~al.}(2001){Santos}, {Israelian}, \&
  {Mayor}}]{Santos-2001}
{Santos}, N.~C., {Israelian}, G., \& {Mayor}, M. 2001, A\&A, 373, 1019

\bibitem[{{Santos} {et~al.}(2004{\natexlab{b}}){Santos}, {Israelian}, \&
  {Mayor}}]{Santos-2004b}
{Santos}, N.~C., {Israelian}, G., \& {Mayor}, M. 2004{\natexlab{b}}, A\&A, 415,
  1153

\bibitem[{{Santos} {et~al.}(2000){Santos}, {Mayor}, {Naef}, {Pepe}, {Queloz},
  {Udry}, \& {Blecha}}]{Santos-2000a}
{Santos}, N.~C., {Mayor}, M., {Naef}, D., {et~al.} 2000, A\&A, 361, 265

\bibitem[{{Sestito} \& {Randich}(2005)}]{Sestito-2005}
{Sestito}, P. \& {Randich}, S. 2005, \aap, 442, 615

\bibitem[{{Sousa} {et~al.}(2008){Sousa}, {Santos}, {Mayor}, {Udry},
  {Casagrande}, {Israelian}, {Pepe}, {Queloz}, \& {Monteiro}}]{Sousa-2008}
{Sousa}, S.~G., {Santos}, N.~C., {Mayor}, M., {et~al.} 2008, A\&A, 487, 373

\bibitem[{{Takeda} \& {Kawanomoto}(2005)}]{Takeda-2005}
{Takeda}, Y. \& {Kawanomoto}, S. 2005, \pasj, 57, 45

\bibitem[{{Takeda} {et~al.}(2007){Takeda}, {Kawanomoto}, {Honda}, {Ando}, \&
  {Sakurai}}]{Takeda-2007}
{Takeda}, Y., {Kawanomoto}, S., {Honda}, S., {Ando}, H., \& {Sakurai}, T. 2007,
  \aap, 468, 663

\bibitem[{{Udry} \& {Santos}(2007)}]{Udry-2007b}
{Udry}, S. \& {Santos}, N.~C. 2007, \araa, 45, 397

\bibitem[{{Valenti} \& {Fischer}(2005)}]{Valenti-2005}
{Valenti}, J.~A. \& {Fischer}, D.~A. 2005, \apjs, 159, 141

\end{thebibliography}

\end{document}